\PassOptionsToPackage{dvipsnames,svgnames,table}{xcolor}

\documentclass{article} 
\usepackage{iclr2025_conference,times}
\usepackage{afterpage}
\iclrfinalcopy

\usepackage{amsmath,amsfonts,bm}

\def\eqref#1{equation~\ref{#1}}

\def\1{\bm{1}}

\DeclareMathAlphabet{\mathsfit}{\encodingdefault}{\sfdefault}{m}{sl}
\SetMathAlphabet{\mathsfit}{bold}{\encodingdefault}{\sfdefault}{bx}{n}

\newcommand{\eg}{\emph{e.g.,}\xspace}

\definecolor{background}{HTML}{FFFFFF}
\definecolor{codebg}{HTML}{FFFFFF}
\definecolor{outputbg}{HTML}{FFFFFF}
\definecolor{textyellow}{HTML}{FFD700}
\definecolor{textred}{HTML}{FF6347}
\definecolor{textgreen}{HTML}{32CD32}
\definecolor{white}{RGB}{255,255,255}

\definecolor{JbBlack}{HTML}{000000}

\definecolor{JbDarkGrey}{RGB}{40, 40, 44}
\definecolor{JbGrey}{HTML}{7D7D7D}
\definecolor{JbLightGrey}{HTML}{CDCDCD}

\definecolor{JbYellow}{HTML}{FCF84A}
\definecolor{JbOrange}{HTML}{FDB60D}
\definecolor{JbOrangeExtra}{HTML}{FC801D}

\definecolor{JbPink}{HTML}{FF318C}
\definecolor{JbDarkPink}{HTML}{DD1265}
\definecolor{JbRed}{HTML}{FE2857}

\definecolor{JbMagenta}{HTML}{FF45ED}
\definecolor{JbPurple}{HTML}{AF1DF5}
\definecolor{JbNavy}{HTML}{6B57FF}

\definecolor{JbBlue}{HTML}{087CFA}
\definecolor{JbLightBlue}{HTML}{07C3F2}

\definecolor{JbGreen}{HTML}{21D789}
\definecolor{JbLightGreen}{HTML}{3DEA62}

\usepackage{tikz}
\pgfdeclarelayer{background}
\pgfdeclarelayer{main}
\pgfsetlayers{background,main}
\usetikzlibrary{calc}

\usepackage{hyperref}
\usepackage{url}
\usepackage{soul}  
\usepackage{xspace}
\usepackage{booktabs}
\usepackage{multirow}
\usepackage{longtable}
\usepackage{graphicx}
\usepackage{array}
\usepackage{tcolorbox}
\usepackage[frozencache,cachedir=.]{minted}
\usepackage{fancyvrb}
\usepackage{array}
\usepackage{tabularx}
\usepackage{booktabs}
\usepackage{makecell}

\title{Themisto: Jupyter-Based Runtime Benchmark}

\author{Konstantin Grotov \& Sergey Titov \\
JetBrains Research\\
\texttt{\{konstantin.grotov,sergey.titov\}@jetbrains.com} \\
}

\begin{document}

\maketitle

\begin{abstract}
In this work, we present a benchmark that consists of Jupyter notebooks development trajectories and allows measuring how large language models (LLMs) can leverage runtime information for predicting code output and code generation. We demonstrate that the current generation of LLMs performs poorly on these tasks and argue that there exists a significantly understudied domain in the development of code-based models, which involves incorporating the runtime context.
\end{abstract}

\section{Introduction}
Recent developments in code completion and generation have been significant. Over the past several years, the field has progressed from generating relatively simple programs~\citep{chen2021evaluating} to solving real-world issues within software repositories~\citep{jimenez2023swe}. However, most studies in this area are based on static snapshots of code~\citep{jiang2024survey}, with only a small body of research exploring the potential of leveraging dynamic code properties, such as runtime information and memory state, for code generation~\citep{chen2024reasoning}. A key reason for this limitation is that common programming environments rarely allow code generation during execution, which is when runtime information can be gathered. 
Jupyter notebooks offer a unique opportunity in this regard—they enable code generation while providing access to runtime information and the current state of the environment. 

In this paper, we present a benchmark designed to measure how models can utilize runtime and environment information, using development trajectories of Jupyter notebooks. A development trajectory is a sequence of Jupyter notebook cell executions in the order performed by a human developer. Each operation includes the cell's content and the runtime state after execution. We propose evaluating a model's ability to predict the code of the next cell to be executed and the output of a given executed cell. 

We believe that by providing these benchmark and baseline results, we can advance the field of incorporating this type of information into code language models. We also make benchmark data available on Zenodo.\footnote{Benchmark data available here: \url{https://zenodo.org/records/14861889}}

\section{Benchmark}

The benchmark consists of a set of Jupyter notebook development trajectories. Each development trajectory consists of all prior cell executions with the given cell's execution context (\eg cell content or runtime snapshot). The order of executions was recorded by the authors of the original dataset (see Section~\ref{data} below) from the notebook development process and is preserved in our benchmark. 
Optionally, for each trajectory one can append the description of the initial task that the notebook was developed for. Using the given trajectory, we evaluate the model's ability to predict the next piece of code to be executed and the output that will be produced by the cell.

\subsection{Tasks and Metrics}

For the benchmark, we have selected two tasks. 

\textbf{Next cell prediction.} In this task, we ask the model to predict the code of the next cell to be executed in our trajectory. This task offers an interesting perspective on code generation, as it requires a significant understanding of the given trajectory to determine what needs to be done next.

\textbf{Cell output prediction.} In this task, we ask the model to predict the output text for the cells with such output type. This task can be challenging for language models in a default snapshot setup, as it requires a strong understanding of the code and effective modeling of the runtime behavior~\citep{gu2024cruxeval}. We hypothesize that providing runtime information should improve scores across the entire set of test trajectories. We suggest that this task demonstrates both the model's capabilities in code modeling and its ability to leverage runtime context.

To measure performance on this task, we use the exact match, ROUGE-L~\citep{lin2004rouge}, and ChrF~\citep{popovic2015chrf} metrics in line with recommendations from~\citet{evtikhiev2023out}.

\subsection{Data}
\label{data}

To acquire the trajectories, we used the JuNE dataset from the paper by~\citet{titov2025june}, where the authors tracked the notebook development process for over 8 hours with a small number of participants. They collected more than 14,000 user events, including
more than 9,000 cell executions during these experiments across 29 notebooks for two original tasks. 
While there are datasets with more notebooks available, such as those from Kaggle~\citep{quaranta2021kgtorrent}, we believe that the JuNE dataset provides more information about the development process. It includes not only the environment and final version of the notebook but also intermediate and debugging steps within the notebook setting, which are the most crucial stages where models should support developers.

To develop our benchmark, we replicated the environment and re-executed four notebooks from the dataset, resulting in a total of 1,453 code executions. Moreover, we collected additional information, such as memory load and execution time of the cell. We also collected and serialized the state of the environment for each step to incorporate it into trajectories. A full list of available context features is given in Table~\ref{tab:dataset_features}. At the end of this process, we obtained the complete trajectory of prior development for each cell execution in the dataset. The example slice of the trajectory content is shown in Figure~\ref{fig:traj_exmaple} and in Appendix~\ref{trajectory-example}.

\begin{table}[h]
    \centering
    \begin{tabular}{p{4cm}p{8cm}}
        \toprule
        \textbf{Feature} & \textbf{Description} \\
        \midrule
        \texttt{kernel\_id} & Unique identifier for the execution kernel \\
        \texttt{code} & Code executed in the cell \\
        \texttt{output} & Output produced by the executed code \\
        \texttt{execution\_time} & Time taken to execute the code in seconds \\
        \texttt{memory\_bytes} & Memory usage during execution in bytes \\
        \texttt{runtime\_variables} & Dictionary of runtime variables in the execution environment. We store each runtime variable's name, size in bytes, and its \texttt{repr} representation. \\
        \texttt{hash\_index} & Unique hash representing the execution state \\
        \bottomrule
    \end{tabular}
    \caption{Descriptions of trajectory step features}
    \label{tab:dataset_features}
\end{table}

The next step involved selecting the cells and outputs to predict. First, we selected all cells with at least five actions in their trajectories. For each task, we filtered out empty examples and extremely long examples, specifically those beyond the 0.99 quantile of the cell length or output length distribution, respectively.

To ensure a diverse set of examples in the benchmark, we selected 200 examples using the following sampling method. First, we randomly chose 180 instances from the second and third quartiles of the output length distribution. Then, we added ten long instances from the fourth quartile and ten short instances from the first quartile.
Additionally, we ensured that no sample in the benchmark to be predicted contains an exception, since the foundational models struggle with stack traces~\citep{gehring2024rlef}. For more information on the statistics of the trajectories and the diversity of examples, please refer to Appendix~\ref{dataset-diversity}.

\begin{figure}[htbp]
    \centering
    \includegraphics[width=0.9\textwidth]{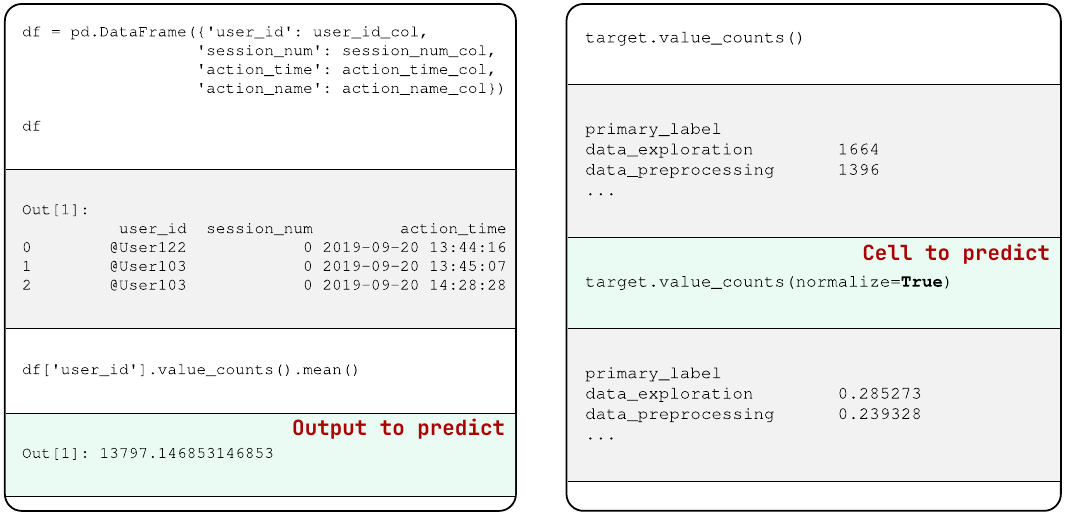}
    \caption{A sample of code-output trajectory pairs for the output prediction task (on the left side) and next cell prediction task (on the right side). The gray and white rows represent the content of the trajectory, including the cell content and the cell output, while the green indicates the entity we aim to predict.}
    \label{fig:traj_exmaple}
\end{figure}

\subsection{Baselines}

To provide an initial baseline for the benchmark, we selected a set of popular language models: GPT-4o~\citep{hurst2024gpt}, GPT-4o-mini~\citep{hurst2024gpt}, Claude 3.5 Sonnet~\citep{anthropic2024claude}, Gemini 1.5 Pro~\citep{team2024gemini}, and DeepSeek-V3~\citep{liu2024deepseek}. We report the benchmark in two settings: using runtime information during inference and without using it.
Additionally, we carried out further post-processing of the model outputs: we removed cell language identifiers and trimmed all redundant spaces and tabulations. Also, we report benchmark results without additional post-processing of model outputs in Appendix~\ref{baselines-raw-results}.
Table~\ref{tab:model-comparison} presents the results for the two tasks of our benchmark, and in Appendix~\ref{baseline-prompts}, you can find details about the inference setup for these models.

\begin{table}[h]
\centering
\begin{tabular}{clcccccc}
\toprule
& \multirow{2}{*}{\textbf{Model}} & \multicolumn{3}{c}{\textbf{Output Prediction}} & \multicolumn{3}{c}{\textbf{Next Cell Prediction}} \\
\cmidrule(lr){3-5} \cmidrule(lr){6-8}
& & Exact Match & RougeL & ChrF & Exact Match & RougeL & ChrF \\
\midrule
\multirow{5}{*}{\makecell{\rotatebox[origin=c]{90}{\textbf{No Runtime}}}} & GPT-4o & 0.16 & 0.32 & 0.47 & 0.10 & 0.28 & 0.39 \\
& GPT-4o-mini & 0.16 & 0.31 & 0.43 & 0.06 & 0.25 & 0.38 \\
& Claude-3.5 & \textbf{0.18} & \textbf{0.38} & 0.50 & 0.12 & 0.30 & 0.42 \\
& Gemini Pro & 0.17 & 0.35 & \textbf{0.54} & 0.12 & 0.34 & 0.43 \\
& DeepSeek-V3 & 0.18 & 0.35 & 0.49 & \textbf{0.13} & \textbf{0.34} & \textbf{0.46} \\
\midrule
\multirow{5}{*}{\makecell{\rotatebox[origin=c]{90}{\textbf{Runtime}}}} & GPT-4o & 0.16 & 0.34 & 0.46 & 0.10 & 0.26 & 0.37 \\
& GPT-4o-mini & 0.15 & 0.30 & 0.43 & 0.07 & 0.27 & 0.36 \\
& Claude-3.5 & 0.09 & 0.34 & 0.48 & 0.11 & 0.30 & 0.42 \\
& Gemini Pro & 0.16 & \textbf{0.35} & \textbf{0.55} & 0.13 & 0.33 & 0.42 \\
& DeepSeek-V3 & \textbf{0.19} & 0.33 & 0.48 & \textbf{0.14} & \textbf{0.35} & \textbf{0.47} \\
\bottomrule
\end{tabular}

\caption{Performance comparison of different foundation models on \textit{Output Prediction} and \textit{Next Cell Prediction} tasks. The metrics shown are Exact Match (higher is better), RougeL F1 score (higher is better), and ChrF score (higher is better).}
\label{tab:model-comparison}
\end{table}

All tested models were able to produce a significant number of exact matches for the output prediction task without leveraging the runtime information. The best results were given by Claude-3.5, with 18\% of cases achieving an exact match. 
All other models achieved very similar results, even though the set of correctly predicted examples differs from model to model. This indicates that the task is equally challenging for different models. 
The models scored higher on Rouge-L and ChrF, suggesting that even in non-exact matches, the models can still produce outputs close to the original. In the setup with runtime information, the scores for the models remained very similar to those without runtime information, except for Claude-3.5, which saw its score drop by half. This indicates that the models are not yet able to effectively leverage the runtime context for this task, highlighting significant potential for further post-training improvements.

The results for next cell prediction show poor overall performance across different types of models, particularly from the perspective of exact match. The best results are produced by DeepSeek-V3, achieving an exact match in 13\% of cases. Similarly to output prediction, these results are compensated by higher scores on ROUGE-L and ChrF, indicating that the models at least produce outputs relevant to the next cell prediction. Despite these numbers, we still consider the results significant—we tasked the models with predicting code for an extremely open-ended task and measured their ability to guess very specific data points, and in some cases, they were able to correctly predict next user actions. 

After runtime inclusion experiments for code predictions, we found that the performance cannot be improved by simply adding all available information in the context and needs to be carefully curated. Although this information is surely valuable for accurate prediction and understanding of the current program state, the actual implementation is an interesting question for the research community. 

\section{Related work}

In recent years, there have been multiple attempts to leverage runtime information in the training process~\citep{ding2024traced, liu2023code, ni2023lever}. Many approaches mainly focus on using the outputs of programs, as seen in recent work by~\citet{gehring2024rlef, dou2024stepcoder, Liu2023RLTFRL}, where they demonstrated that models poorly respond to compiler feedback and suggested a reinforcement learning approach to improve code generation results in a multi-shot setup. However, other works, such as TRACED~\citep{ding2024traced}, show that the addition of runtime information can improve the behavior of the model to predict execution states or locate bugs.

There are two notable benchmarks that assess a model's ability to simulate code execution: CruxEval~\citep{gu2024cruxeval} and REval~\citep{chen2024reasoning}. CruxEval proposes triplets of code, input, and output, asking models to predict the input or output given the other two. They demonstrate that a chain of thought setup is more efficient, highlighting that reasoning plays a key role in modeling the execution process. REval builds on CruxEval by adding runtime data to the test set and introduces novel tasks like program state prediction and execution path prediction, in addition to output prediction. They show that reasoning capabilities vary significantly among models. For example, in execution path prediction, even the strongest tested model, GPT-4-Turbo, only achieves an accuracy of 57.7\%.


\section{Threat to validity and Conclusion}

We believe that our benchmark can provide strong momentum toward the utilization of runtime information in code-based models. We hope that the unique environment of Jupyter notebooks can leverage newly fine-tuned models, making the notebook development process more pleasant and productive. However, the current version of the benchmark has several significant issues.

The main problem is low variability. The original data was collected from only two tasks and 20 participants, and we used only a subsample of this data. This leads to a severely underrepresented space of notebook trajectories. This limitation makes it difficult to draw conclusions about the generalizability of approaches that work with runtime context. Given the community's interest in the benchmark, one may use the tooling provided in the original JuNE dataset to gather more data for additional tasks.

With this benchmark, we introduce a new dynamic modality for code generation and program analysis, moving beyond static code base snapshots to incorporate complete development trajectories. This approach makes runtime information and development progress available to models, potentially allowing them to better align with developers' workflows and expectations. Our findings demonstrate that this is a challenging problem that remains difficult even for advanced foundation models, opening new horizons for future research in areas such as runtime-aware code completion, dynamic context understanding, and interactive development assistance.

\bibliography{iclr2025_conference}

\begin{thebibliography}{20}
\providecommand{\natexlab}[1]{#1}
\providecommand{\url}[1]{\texttt{#1}}
\expandafter\ifx\csname urlstyle\endcsname\relax
  \providecommand{\doi}[1]{doi: #1}\else
  \providecommand{\doi}{doi: \begingroup \urlstyle{rm}\Url}\fi

\bibitem[Anthropic(2024)]{anthropic2024claude}
AI~Anthropic.
\newblock Claude 3.5 sonnet model card addendum.
\newblock \emph{Claude-3.5 Model Card}, 3:\penalty0 1--8, 2024.

\bibitem[Chen et~al.(2024)Chen, Pan, Hu, Li, Li, and Xia]{chen2024reasoning}
Junkai Chen, Zhiyuan Pan, Xing Hu, Zhenhao Li, Ge~Li, and Xin Xia.
\newblock Reasoning runtime behavior of a program with llm: How far are we?
\newblock \emph{arXiv preprint cs.SE/2403.16437}, 2024.

\bibitem[Chen et~al.(2021)Chen, Tworek, Jun, Yuan, Pinto, Kaplan, Edwards, Burda, Joseph, Brockman, et~al.]{chen2021evaluating}
Mark Chen, Jerry Tworek, Heewoo Jun, Qiming Yuan, Henrique Ponde De~Oliveira Pinto, Jared Kaplan, Harri Edwards, Yuri Burda, Nicholas Joseph, Greg Brockman, et~al.
\newblock Evaluating large language models trained on code.
\newblock \emph{arXiv preprint arXiv:2107.03374}, 2021.

\bibitem[Ding et~al.(2024)Ding, Steenhoek, Pei, Kaiser, Le, and Ray]{ding2024traced}
Yangruibo Ding, Benjamin Steenhoek, Kexin Pei, Gail Kaiser, Wei Le, and Baishakhi Ray.
\newblock Traced: Execution-aware pre-training for source code.
\newblock In \emph{Proceedings of the 46th IEEE/ACM International Conference on Software Engineering}, pp.\  1--12, 2024.

\bibitem[Dou et~al.(2024)Dou, Liu, Jia, Xiong, Zhou, Shen, Shan, Huang, Wang, Fan, et~al.]{dou2024stepcoder}
Shihan Dou, Yan Liu, Haoxiang Jia, Limao Xiong, Enyu Zhou, Wei Shen, Junjie Shan, Caishuang Huang, Xiao Wang, Xiaoran Fan, et~al.
\newblock Stepcoder: Improve code generation with reinforcement learning from compiler feedback.
\newblock \emph{arXiv preprint arXiv:2402.01391}, 2024.

\bibitem[Evtikhiev et~al.(2023)Evtikhiev, Bogomolov, Sokolov, and Bryksin]{evtikhiev2023out}
Mikhail Evtikhiev, Egor Bogomolov, Yaroslav Sokolov, and Timofey Bryksin.
\newblock Out of the bleu: how should we assess quality of the code generation models?
\newblock \emph{Journal of Systems and Software}, 203:\penalty0 111741, 2023.

\bibitem[Gehring et~al.(2024)Gehring, Zheng, Copet, Mella, Cohen, and Synnaeve]{gehring2024rlef}
Jonas Gehring, Kunhao Zheng, Jade Copet, Vegard Mella, Taco Cohen, and Gabriel Synnaeve.
\newblock Rlef: Grounding code llms in execution feedback with reinforcement learning.
\newblock \emph{arXiv preprint arXiv:2410.02089}, 2024.

\bibitem[Gu et~al.(2024)Gu, Rozi{\`e}re, Leather, Solar-Lezama, Synnaeve, and Wang]{gu2024cruxeval}
Alex Gu, Baptiste Rozi{\`e}re, Hugh Leather, Armando Solar-Lezama, Gabriel Synnaeve, and Sida~I Wang.
\newblock Cruxeval: A benchmark for code reasoning, understanding and execution.
\newblock \emph{arXiv preprint arXiv:2401.03065}, 2024.

\bibitem[Hurst et~al.(2024)Hurst, Lerer, Goucher, Perelman, Ramesh, Clark, Ostrow, Welihinda, Hayes, Radford, et~al.]{hurst2024gpt}
Aaron Hurst, Adam Lerer, Adam~P Goucher, Adam Perelman, Aditya Ramesh, Aidan Clark, AJ~Ostrow, Akila Welihinda, Alan Hayes, Alec Radford, et~al.
\newblock Gpt-4o system card.
\newblock \emph{arXiv preprint arXiv:2410.21276}, 2024.

\bibitem[Jiang et~al.(2024)Jiang, Wang, Shen, Kim, and Kim]{jiang2024survey}
Juyong Jiang, Fan Wang, Jiasi Shen, Sungju Kim, and Sunghun Kim.
\newblock A survey on large language models for code generation.
\newblock \emph{arXiv preprint arXiv:2406.00515}, 2024.

\bibitem[Jimenez et~al.(2023)Jimenez, Yang, Wettig, Yao, Pei, Press, and Narasimhan]{jimenez2023swe}
Carlos~E Jimenez, John Yang, Alexander Wettig, Shunyu Yao, Kexin Pei, Ofir Press, and Karthik Narasimhan.
\newblock Swe-bench: Can language models resolve real-world github issues?
\newblock \emph{arXiv preprint arXiv:2310.06770}, 2023.

\bibitem[Lin(2004)]{lin2004rouge}
Chin-Yew Lin.
\newblock Rouge: A package for automatic evaluation of summaries.
\newblock In \emph{Text summarization branches out}, pp.\  74--81, 2004.

\bibitem[Liu et~al.(2024)Liu, Feng, Xue, Wang, Wu, Lu, Zhao, Deng, Zhang, Ruan, et~al.]{liu2024deepseek}
Aixin Liu, Bei Feng, Bing Xue, Bingxuan Wang, Bochao Wu, Chengda Lu, Chenggang Zhao, Chengqi Deng, Chenyu Zhang, Chong Ruan, et~al.
\newblock Deepseek-v3 technical report.
\newblock \emph{arXiv preprint arXiv:2412.19437}, 2024.

\bibitem[Liu et~al.(2023{\natexlab{a}})Liu, Lu, Chen, Jiang, Svyatkovskiy, Fu, Sundaresan, and Duan]{liu2023code}
Chenxiao Liu, Shuai Lu, Weizhu Chen, Daxin Jiang, Alexey Svyatkovskiy, Shengyu Fu, Neel Sundaresan, and Nan Duan.
\newblock Code execution with pre-trained language models.
\newblock \emph{arXiv preprint arXiv:2305.05383}, 2023{\natexlab{a}}.

\bibitem[Liu et~al.(2023{\natexlab{b}})Liu, Zhu, Xiao, Fu, Han, Yang, and Ye]{Liu2023RLTFRL}
Jiate Liu, Yiqin Zhu, Kaiwen Xiao, Qiang Fu, Xiao Han, Wei Yang, and Deheng Ye.
\newblock Rltf: Reinforcement learning from unit test feedback.
\newblock \emph{Trans. Mach. Learn. Res.}, 2023, 2023{\natexlab{b}}.
\newblock URL \url{https://api.semanticscholar.org/CorpusID:259501019}.

\bibitem[Ni et~al.(2023)Ni, Iyer, Radev, Stoyanov, Yih, Wang, and Lin]{ni2023lever}
Ansong Ni, Srini Iyer, Dragomir Radev, Veselin Stoyanov, Wen-tau Yih, Sida Wang, and Xi~Victoria Lin.
\newblock Lever: Learning to verify language-to-code generation with execution.
\newblock In \emph{International Conference on Machine Learning}, pp.\  26106--26128. PMLR, 2023.

\bibitem[Popovi{\'c}(2015)]{popovic2015chrf}
Maja Popovi{\'c}.
\newblock chrf: character n-gram f-score for automatic mt evaluation.
\newblock In \emph{Proceedings of the tenth workshop on statistical machine translation}, pp.\  392--395, 2015.

\bibitem[Quaranta et~al.(2021)Quaranta, Calefato, and Lanubile]{quaranta2021kgtorrent}
Luigi Quaranta, Fabio Calefato, and Filippo Lanubile.
\newblock Kgtorrent: A dataset of python jupyter notebooks from kaggle.
\newblock In \emph{2021 IEEE/ACM 18th International Conference on Mining Software Repositories (MSR)}, pp.\  550--554. IEEE, 2021.

\bibitem[Team et~al.(2024)Team, Georgiev, Lei, Burnell, Bai, Gulati, Tanzer, Vincent, Pan, Wang, et~al.]{team2024gemini}
Gemini Team, Petko Georgiev, Ving~Ian Lei, Ryan Burnell, Libin Bai, Anmol Gulati, Garrett Tanzer, Damien Vincent, Zhufeng Pan, Shibo Wang, et~al.
\newblock Gemini 1.5: Unlocking multimodal understanding across millions of tokens of context.
\newblock \emph{arXiv preprint arXiv:2403.05530}, 2024.

\bibitem[Titov et~al.(2025)Titov, Grotov, Sarasua, Golubev, Ramasamy, Bacchelli, Bernstein, and Bryksin]{titov2025june}
Sergey Titov, Konstantin Grotov, Cristina Sarasua, Yaroslav Golubev, Dhivyabharathi Ramasamy, Alberto Bacchelli, Abraham Bernstein, and Timofey Bryksin.
\newblock June: Jupyter notebooks executions dataset.
\newblock \url{https://huggingface.co/datasets/JetBrains-Research/JuNE}, 2025.
\newblock Dataset containing logs of code evolution in Jupyter notebooks, comprising over 100 hours of execution logs from 20 participants solving data science tasks.

\end{thebibliography}
\bibliographystyle{iclr2025_conference}
\appendix
\section{Appendix} \label{appendix}

\subsection{Diversity of the Samples in Benchmark} \label{dataset-diversity}

\begin{figure}[htbp]
    \centering
    \includegraphics[width=0.9\textwidth]{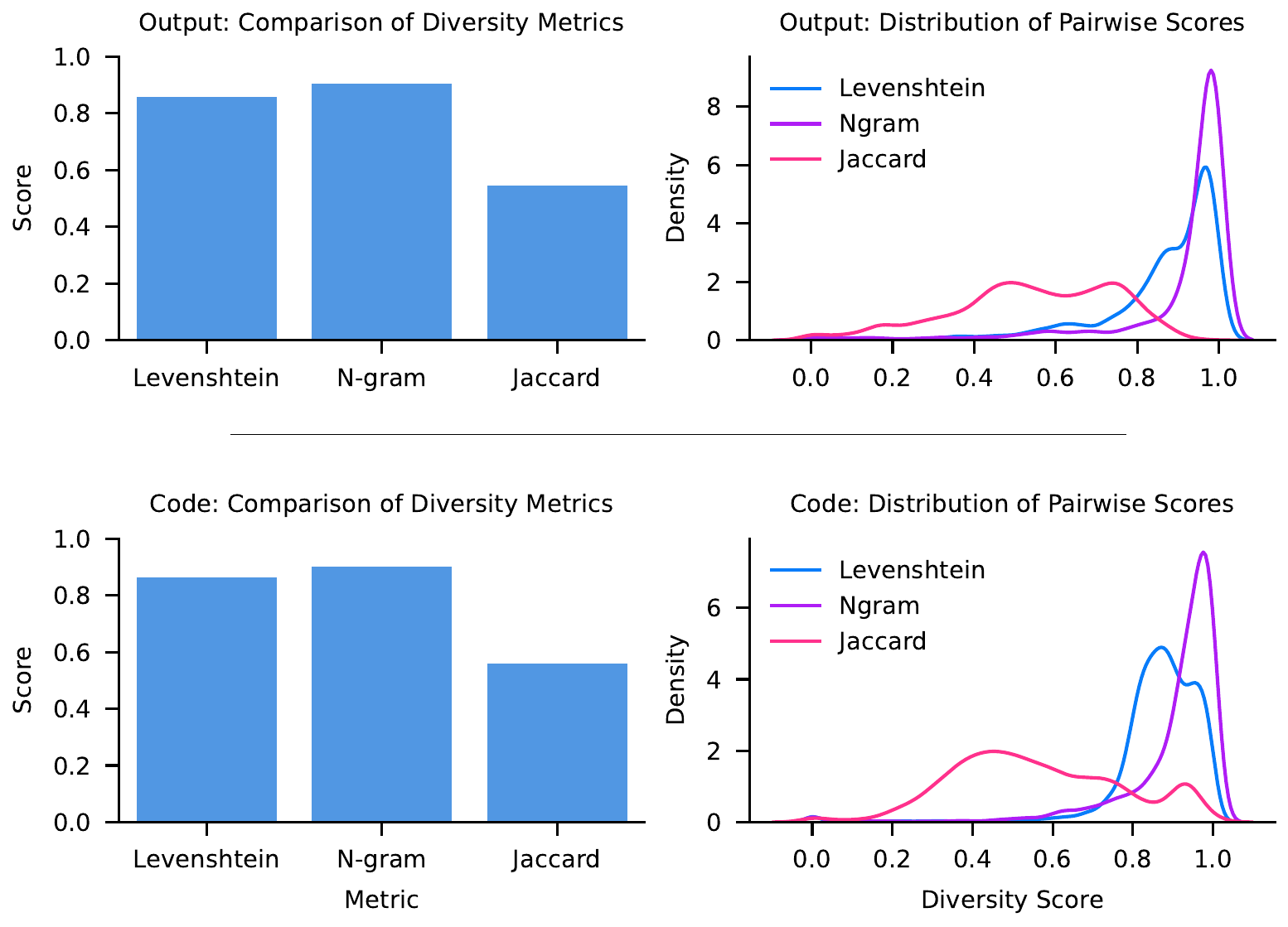}
    \caption{Diversity metrics comparison between output and code.}
    \label{fig:diversity-plots}
\end{figure}

\subsection{Inference Setup} \label{baseline-prompts}

\begin{tcolorbox}[fonttitle=\bfseries\color{white}, colback=outputbg, colframe=gray, boxsep=1mm, left=2mm, right=2mm, top=1mm, bottom=1mm, title=Output Prediction]
\begin{minted}[fontsize=\tiny, style=bw]{text}
You are a Python REPL interpreter. Given a sequence of executed Python code cells and their outputs,
predict the output of the next executed code cell. Provide only the output, exactly as it would appear
in a Python interpreter. YOU MUST NOT include any additional tags (```python, ```, etc).
Previous code cells and their outputs:

Code: {{code}}
Output: {{output}}
...
Code: {{code}}
Output: {{output}}

Predict the output for this code: {{code_to_predict}}
\end{minted}
\end{tcolorbox}
\begin{tcolorbox}[fonttitle=\bfseries\color{white}, colback=outputbg, colframe=gray, boxsep=1mm, left=2mm, right=2mm, top=1mm, bottom=1mm, title=Next Cell Prediction]
\begin{minted}[fontsize=\tiny, style=bw]{text}
You are an expert Python programmer. Given a sequence of executed Python code cells and their outputs,
predict what the next code cell will be. Provide only the code, exactly as it would be written in a 
Jupyter notebook. YOU MUST NOT include any additional tags (```python, ```, etc).
Previous code cells and their outputs:

Code: {{code}}
Output: {{output}}
...
Code: {{code}}
Output: {{output}}

Predict the next code cell that would logically follow:
\end{minted}
\end{tcolorbox}

\subsection{Baselines Without Output Processing}
\label{baselines-raw-results}

\begin{table}[h]
\centering
\begin{tabular}{lcccccc}
\toprule
\multirow{2}{*}{\textbf{Model}} & \multicolumn{3}{c}{\textbf{Output Prediction}} & \multicolumn{3}{c}{\textbf{Next Cell Prediction}} \\
\cmidrule(lr){2-4} \cmidrule(lr){5-7}
& Exact Match & RougeL & ChrF & Exact Match & RougeL & ChrF \\
\midrule
GPT-4o$^{\text{Runtime}}$ & \textbf{0.10} & 0.53 & 0.52 & 0 & 0.17 & 0.26 \\
GPT-4o-mini$^{\text{Runtime}}$ & 0.08 & 0.37 & 0.41 & 0 & 0.16 & 0.23 \\
\midrule
GPT-4o & \textbf{0.10} & 0.54 & 0.55 & 0 & 0.23 & 0.31 \\
GPT-4o-mini & 0.09 & 0.51 & 0.51 & 0 & 0.21 & 0.31 \\
Claude-3.5 & 0.06 & 0.50 & 0.52 & 0.01 & 0.10 & 0.21 \\
Gemini 1.5 Pro & 0.05 & \textbf{0.55} & \textbf{0.57} & 0.02 & 0.27 & 0.32 \\
DeepSeek-V3 & 0.08 & 0.53 & 0.56 & \textbf{0.07} & \textbf{0.34} & \textbf{0.40} \\

\bottomrule
\end{tabular}
\caption{Performance comparison of different foundation models on \textit{Output Prediction} and \textit{Next Cell Prediction} tasks \textbf{without additional output processing}. The metrics shown are Exact Match (higher is better), RougeL F1 score (higher is better), and ChrF score (higher is better).}
\label{tab:model-comparison}
\end{table}

\subsection{Example of Trajectory}
\label{trajectory-example}

\begin{tcolorbox}[colback=background, colframe=JbNavy, title=Step 70, 
  fonttitle=\bfseries\color{white}, boxsep=1mm, left=2mm, right=2mm, top=1mm, bottom=1mm]
\begin{tcolorbox}[colback=codebg, colframe=gray, boxsep=1mm, left=2mm, right=2mm, top=1mm, bottom=1mm, title=Code:]
\begin{minted}[fontsize=\tiny]{python}
df
\end{minted}
\end{tcolorbox}

\begin{tcolorbox}[colback=background, colframe=gray, boxsep=1mm, left=2mm, right=2mm, top=1mm, bottom=1mm, title=Runtime Variables:]

\begin{minted}[fontsize=\tiny]{python}
action: {'size': 80, 'type': 'str', 'value': 'Action_7 (27/06...)'}
...
action_time: {'size': 68, 'type': 'str', 'value': '27/06/20 | 17:3...'}
df: {'size': 79714318, 'type': 'DataFrame', 'value': '       user_id ...'}
\end{minted}
\end{tcolorbox}

\begin{tcolorbox}[colback=outputbg, colframe=gray, boxsep=1mm, left=2mm, right=2mm, top=1mm, bottom=1mm, title=Output:]
\begin{minted}[fontsize=\tiny]{python}
   user_id                                               info
0   User92  Action_3 (15/10/19 | 18:08:02) -> Action_1 (15...
1  User140  Action_3 (15/05/20 | 15:37:04) -> Action_8 (15...
2  User105  Action_4 (25/04/20 | 01:08:29) -> Action_7 (25...
...
[87192 rows x 2 columns]
\end{minted}
\end{tcolorbox}

Execution Time: 0.01 seconds \\
Memory Usage: 10250.90 MB
\end{tcolorbox}


\begin{tcolorbox}[colback=background, colframe=JbDarkPink, title=Step 71 (to predict), 
  fonttitle=\bfseries\color{white}, boxsep=1mm, left=2mm, right=2mm, top=1mm, bottom=1mm]

\begin{tcolorbox}[colback=codebg, colframe=gray, boxsep=1mm, left=2mm, right=2mm, top=1mm, bottom=1mm, title=Code:]
\begin{minted}[fontsize=\tiny]{python}
df = df.assign(actions=df['info'].str.split('-> ')).explode('actions')
df
\end{minted}
\end{tcolorbox}

\begin{tcolorbox}[colback=background, colframe=gray, boxsep=1mm, left=2mm, right=2mm, top=1mm, bottom=1mm, title=Runtime Variables:]
\begin{minted}[fontsize=\tiny]{python}
action: {'size': 80, 'type': 'str', 'value': 'Action_7 (27/06...)'}
...
action_time: {'size': 68, 'type': 'str', 'value': '27/06/20 | 17:3...'}
df: {'size': 75310105, 'type': 'DataFrame', 'value': '       user_id ...'}
\end{minted}
\end{tcolorbox}

\begin{tcolorbox}[colback=outputbg, colframe=gray, boxsep=1mm, left=2mm, right=2mm, top=1mm, bottom=1mm, title=Ground Truth Output:]
\begin{minted}[fontsize=\tiny]{python}
   user_id  ...                           actions
0   User92  ...   Action_3 (15/10/19 | 18:08:02)
0   User92  ...   Action_1 (15/10/19 | 18:54:49)
0   User92  ...  Action_10 (15/10/19 | 20:02:54)
...
[2053521 rows x 3 columns]
\end{minted}
\end{tcolorbox}
\end{tcolorbox}

\end{document}